\newcommand{\be}{\begin{equation}}\newcommand{\ee}{\end{equation}}
\newcommand{\bea}{\begin{eqnarray}}\newcommand{\eea}{\end{eqnarray}}
\newcommand{\p}[1]{(\ref{#1})}

\newcommand\s{\scriptscriptstyle}
\newcommand{\E}{\stackrel{\to}{E}}
\newcommand{\M}{\stackrel{\to}{M}}

\documentstyle[12pt]{article}
\begin{document}
\renewcommand{\thefootnote}{\fnsymbol{footnote}}
\begin{flushright}
hep-th/9801016 \\
January, 1998
\end{flushright}
\begin{center}
{\large\bf Modifying N=2 Supersymmetry via Partial Breaking}
\vspace{0.6cm} \\
{\large Evgeny Ivanov, Boris Zupnik}
\footnote{On leave of absence from: Institute of Applied Physics,
Tashkent State University, Uzbekistan}
\vspace{0.4cm} \\
{\it Bogoliubov Laboratory of Theoretical Physics, JINR,  \\
141980 Dubna, Moscow Region, Russia} 

\vspace{0.4cm}
{\large \it Talk given at the 31th International Symposium Ahrenshoop \\
on the Theory of Elementary Particles, \\
Buckow (Germany), September 2 - 6, 1997}
\end{center}

\vspace{0.3cm}
\begin{abstract}
\noindent{\small 
We study realization of $N=2$ SUSY in $N=2$
abelian gauge theory with electric and magnetic $FI$ terms 
within a manifestly supersymmetric formulation. 
We find that after dualization of even one 
$FI$ term $N=2$ SUSY is realized in a partial breaking mode off shell. 
In the case of two $FI$ terms, this regime is preserved on shell. 
The $N=2$ SUSY algebra is shown to be modified on gauge-variant 
objects.} 
\end{abstract}
\renewcommand{\thefootnote}{\arabic{footnote}}
\setcounter{footnote}0

\vspace{0.2cm}
\noindent{\bf 1.} A celebrated mechanism of spontaneous 
breakdown of rigid $N=2$ SUSY consists in adding 
a Fayet-Iliopoulos ($FI$) term to the
action of $N=2$ gauge theory. Recently, Antoniadis, Partouche and Taylor
(APT) \cite{APT} have found that the dual formulation of $N=2$ abelian
gauge theory \cite{SW}
provides a
more general framework for such a spontaneous breaking due to the
possibility to define two kinds of the $FI$ terms (see also \cite{Fe}).
One of them (`electric') is standard, while another (`magnetic') is
related to a dual $U(1)$ gauge supermultiplet. APT show that a partial
spontaneous breakdown of $N=2$ SUSY to $N=1$ becomes possible,
if one starts with an effective $N=2$ Maxwell action 
and simultaneously includes two such $FI$ terms.

Here we report on the results of studying 
the invariance properties of $N=2$ Maxwell 
action with the two types of $FI$-terms in a 
manifestly supersymmetric $N=2$ superfield approach. 
Our basic observation is that
after duality transformation of a system even with one sort of the $FI$
term, off-shell $N=2$ SUSY starts to be realized in a {\it partial} 
spontaneous breaking mode. 
When {\it both} $FI$ terms are included, 
this partial breaking 
is retained on shell by the APT mechanism. 
We study how these modified $N=2$ SUSY transformations act on the 
gauge potentials and find that the 
$N=2$ SUSY algebra also undergoes a modification. 

\vspace{0.2cm}
\noindent {\bf 2.} Let us start from the following  
representation of the superfield action of $N=2$ Maxwell theory 
with the $FI$ term \cite{IZ}
\bea
S(W,L) &=& \frac{i}{4}\int d^4 x d^4 \theta\; [{\cal F}(W) 
- WW_L  +\frac{i}{2} E^{ik} (\theta_i\theta_k)W] + \mbox{c.c.} 
\nonumber \\
&\equiv & S(W) + S_L + S_e~, \qquad 
W_L = 
(\bar{D})^4 D_{ik} L^{ik} \;.\label{1}
\eea
Here, ${\cal F}(W)$ is an arbitrary holomorphic function, $W$ is a chiral 
$N=2$ superfield, $L^{ik}$ is a real unconstrained 
$N=2$ superfield Lagrange multiplier, $E^{ik}$ is a real 
$SU(2)$ triplet of constants and $D^{ik}=D^{i\alpha} D^k_{\alpha} $. 
Varying $L^{ik}$ yields the  
constraint \cite{GSW}
\be
D^{ik} W - \bar{D}^{ik} \overline{W} = 0 \label{A7}
\ee
which can be solved in terms of Mezincescu prepotential $V^{ik}$ 
\cite{Me}
\be
W \equiv W_{\s V} = (\bar{D})^4 D_{ik} V^{ik}~.
\label{A8}
\ee
Upon substituting this solution back into \p{1}, the latter becomes 
\be
S(W,L) \Rightarrow S(V) = S(W_{\s V}) + S_e\;, \quad 
S_e = \int d^4 x d^4 \theta d^4 \bar \theta\; E_{ik}V^{ik} 
\label{2} 
\ee
that is the standard `electric' form of the action of $N=2$ Maxwell 
theory with $FI$ term. On the other hand, 
one can first vary \p{1} with respect to $W$, which yields  
\be
\frac{\partial {\cal F}}{\partial W} = W_L -(i/2)(\theta_k\theta_l)E^{kl}
\equiv \hat{W}_L 
\;, \label{hatW}
\ee
\be
S(W,L) \Rightarrow 
\frac{i}{4}\int d^4 x d^4 \theta \;\hat{{\cal F}}(\hat{W}_L) + 
\mbox{c.c.}\;, 
\;\; \hat{{\cal F}}(\hat{W_L}) \equiv  {\cal F}(W(\hat{W_L})) - 
\hat{W}_L\cdot 
W(\hat{W}_L)\;. \label{SL}
\ee
In this dual, `magnetic'  representation the $FI$ term-modified 
$N=2$ Maxwell action is expressed through the 
dual (`magnetic') superfield strength and prepotential $W_L$ and $L^{ik}$. 
Thus, \p{1} is 
a sort of `master' action from which both the electric and magnetic 
forms of the $N=2$ Maxwell action can be obtained (see \cite{SW,He} for 
a similar discussion of the standard $E^{ik} = 0$ case).

Let us discuss peculiarities of realization of $N=2$ SUSY in the 
magnetic representation. The electric action \p{2} 
is invariant under the standard $N=2$ SUSY. The same is true for the 
dual actions \p{1}, \p{SL} in the absence of the $FI$ term. 
However, if $E^{ik}\neq 0$, the conventional $N=2$ 
SUSY gets broken in \p{1}, \p{SL}. The only way to 
restore it 
is to modify the 
transformation law of $W_L$ 
\be
\delta_\epsilon W_{\s L} = i(\epsilon_k\theta_l)E^{kl} +
 i(\epsilon Q + \bar{\epsilon}\bar{Q})W_{\s L}\;,
 \label{B9}
\ee
where $Q^i_\alpha, \bar Q^i_{\dot\alpha}$ are standard generators of 
$N=2$ SUSY. It is easy to find the appropriate modified 
transformation of $L^{ik}$. 

An inhomogeneous term in the $N=2$ SUSY 
transformation law means spontaneous breakdown of 
$N=2$ SUSY. To see which kind of breaking occurs in the case at hand, 
let us firstly discuss the standard electric action \p{2}. 

Spontaneous breaking of $N=2$ SUSY 
by the $FI$ term is related to the possibility of 
a non-zero vacuum solution for the auxiliary component 
$X^{ik}  = -{1\over 4}D^{ik}W|_0$ 
\be
<X^{ik}>\equiv x^{ik} \sim\;E^{ik}~. \label{realaux}
\ee
Provided that such a solution of the equations of motion exists 
and corresponds to a stable
classical vacuum, there appears an inhomogeneous term in the
on-shell SUSY transformation law of the $N=2$ gaugini
doublet $\lambda^{i\alpha}$
\be  \label{goldstin}
\delta \lambda^{i\alpha} \sim \epsilon^\alpha_k E^{ik}\;,
\ee
$\epsilon^{\alpha}_k$ being the transformation parameter.
Thus there are Goldstone fermions in the theory, which is a standard
signal of spontaneous breaking of $N=2$ SUSY. Since the 
inhomogeneous terms in \p{goldstin} appear as a result of solving 
equations of motion, it is natural to call $\lambda^{i\alpha}$
{\it on-shell} Goldstone fermions. As the matrix $E^{ik}$ is 
non-degenerate, {\it both} $\lambda^{1\alpha}, \lambda^{2\alpha}$ 
are shifted by independent parameters, and so they both are 
Goldstone fermions. 
Thus, with the standard electric $FI$ term, only {\it total} 
spontaneous breaking of $N=2$ SUSY can occur (actually, for one 
vector multiplet this is possible only in the free case, 
with quadratic function ${\cal F}$ \cite{GO}).

In the dual, magnetic representation of the same theory corresponding to 
the actions \p{1} or \p{SL} the situation is radically 
different: the off-shell transformation law \p{B9} contains 
an `inborn' inhomogeneous piece. This leads us to interpret 
the magnetic gaugini as  {\it off-shell} Goldstone fermions.  

Both gaugini are shifted in \p{B9}, so at first sight we 
are facing the total
off-shell spontaneous breaking of $N=2$ SUSY in this case. 
However, by a proper shift 
\be
W_{\s L} \rightarrow \widetilde{W}_{\s L} = W_{\s L} +
{1\over 2}(\theta_i\theta_k)C^{ik},         \label{shiftWL}
\ee
one can restore a homogeneous transformation law with respect to
{\it one} of two $N=1$ supersymmetries contained in $N=2$
SUSY (it is easy to find the appropriate redefinition of
$L^{ik}$). The object $\widetilde{W}_{\s L}$ transforms
as follows
\be
\delta_\epsilon \widetilde{W}_{\s L} = (\epsilon_k\theta_l)(C^{kl} 
+iE^{kl})+ i(\epsilon Q + \bar{\epsilon}\bar{Q})\widetilde{W}_{\s L}\;.
\label{B9til}
\ee
One can always choose $C^{ik}$ so that
\be  \label{degener1}
\mbox{det}\;(C+iE) = 0~.
\ee
This means that $C^{ik} + iE^{ik}$ is a degenerate
symmetric $2\times 2$ matrix, so it can be brought to the form with only 
one non-zero entry. As a result, $\widetilde{W}_{\s L}$ is actually 
shifted under the action of only {\it one} linear combination of the
modified $N=2$ SUSY generators $\hat{Q}^{1,2}_\alpha$. 
The same is true for the physical fermionic components:
only {\it one} their combination is the genuine off-shell Goldstone
fermion. 

Thus we arrive at the important conclusion: in the dual, magnetic
representation of $N=2$ Maxwell theory with $FI$ term $N=2$ 
SUSY is realized {\it off shell} in a {\it
partial spontaneous breaking mode}, so that some $N=1$ SUSY
remains unbroken. 

It is straightforward 
to show that the action \p{SL} leads to the same vacuum structure as 
the original electric action \p{2}.
Thus on shell in the magnetic representation
we again encounter the total spontaneous breaking of $N=2$ SUSY.

\vspace{0.2cm}
\noindent {\bf 3.} Let us show that the phenomenon of partial breaking 
of $N=2$ SUSY becomes valid both off and on shell upon adding to the 
`master' action \p{1} the new sort of $FI$ term, the `magnetic' one: 
\be
S(W,L) \Rightarrow S(W,L)' = S(W,L) + S_m \label{SWLm}~, \quad 
S_{m}=\frac{1}{8}\int d^{4}x d^4\theta\; M^{kl} 
(\theta_k\theta_l)W_{\s L}
+ \mbox{c.c.}~,
\label{B13}
\ee
$M^{ik}$ being another triplet of real constants.
It is easy to show that $S_{m}$ is
invariant under the Goldstone-type transformation (\ref{B9}). 

When one descends to the electric representation of 
\p{B13} (by varying $L^{ik}$),
the only effect of the magnetic $FI$ term $S_m$ is the modification of the
constraint \p{A7}: 
\be
D^{ik}W - \bar D^{ik}\overline{W} = 4iM^{ik}~  .
\label{B14}
\ee
It suggests the redefinition
\be \label{B14b}
W = W_{\s V} - {i\over 2}(\theta_i\theta_k)M^{ik}~,
\ee
with $W_{\s V}$ satisfying eq. \p{A7} and hence given
by eq. \p{A8}. 
This shift amounts to the appearance
of the constant imaginary part $-(i/2)M^{ik}$ in the auxiliary field of
$W$,
$$
\hat{X}^{ik} \equiv -{1\over 4} D^{ik} W|_0 = X^{ik} - {i\over 2}M^{ik}~.
$$

The inclusion of magnetic $FI$ term cannot change the standard 
transformation properties of $W$ under $N=2$ SUSY. Then the 
relation \p{B14b} requires to modify the transformation
law of $W_{\s V}$, and, respectively, of $V^{ik}$ on the pattern of eq. 
\p{B9}
\be
\delta_\epsilon W_{\s V} = i(\epsilon_k\theta_l)M^{kl} +
 i(\epsilon Q + \bar{\epsilon}\bar{Q})W_{\s V}
\label{modtraV}
\ee
(it is easy to find the appropriate transformation law of $V^{ik}$). 
In other words, $N=2$ SUSY
is now realized in a Goldstone-type fashion in the electric
representation as well, but with $M^{ik}$ instead of $E^{ik}$ as the
`structure' constants. 
So, when
both $FI$ terms are present, 
there is no way to restore the standard $N=2$ SUSY off shell.
The same arguments as in the previous
Section show that $N=2$ SUSY in both representations is realized
off shell in {\it the partial breaking mode}.

A general electric effective action of the abelian gauge model with the
$(E,M)$- mechanism of the spontaneous breaking can be obtained by
substituting the expression for $W$, eq.\p{B14b}, into the action \p{SWLm}:
\be
S_{\s (E,M)}=\left [\frac{i}{4}\int d^4x d^4\theta \;{\cal F}(W)
 + \mbox{c.c.} \right ] + \int d^4 x d^4 \theta d^4 \bar \theta 
\; E_{ik} V^{ik}~.
\label{B14c}
\ee
Taking the standard vacuum ansatz 
\be  \label{anz1}
<W_{\s V}>_{\s 0} = a + (\theta_i\theta_k)\; x^{ik}\;,
\ee
it is easy to show that the superfield equation of motion following from 
\p{B14c} implies the following equations for moduli $a$, $x^{ik}$
\be 
(i)\;\;x^{kl} = \frac{1}{2\tau_2(a)}\left( \tau_1(a) M^{kl} - 
E^{kl} \right)~,
\quad (ii)\;\; \tau^\prime\;\hat{x}^{ik} \hat{x}_{ik} = 0~, 
\label{IIequ}
\ee
where 
$$
\tau = {\cal F}'' = \tau_1 + i\tau_2~, \qquad 
\hat{x}^{ik} = <\hat{X}^{ik}> = x^{ik} - (i/2)M^{ik}~.
$$
A crucial new point compared to the case of $M^{ik} = 0$ 
is that the vector
$\hat{x}^{ik} = x^{ik} - (i/2)M^{ik}$ is {\it complex}, so the vanishing 
of its square does not imply it to vanish. 
As a result, besides the trivial solution
$\tau^\prime = 0$ \cite{GO}, eq.{\it (ii)} in \p{IIequ} possesses 
the non-trivial one
\be \label{degener}
\tau^\prime \neq 0~, \qquad \hat{x}^{ik}\hat{x}_{ik} = 0~.
\ee
This solution amounts
to the following relations 
\be
\tau_1(a)=\frac{\E\M}{\M^2}~,\quad |\tau_2(a)|=
\frac{\sqrt{\E^2\M^2-(\E\M)^2}}{\M^2} = \frac{|\E \times \M|}{\M^2}~.
\label{taurestr}
\ee
This is just the vacuum solution found in \cite{APT} based on 
the component version of the
action \p{B14c}. It triggers a {\it partial} spontaneous breaking of 
$N=2$ SUSY down to $N=1$. Only one 
combination of gaugini is the Goldstone fermion in this case.

Thus, the phenomenon of the off-shell partial breaking of $N=2$ SUSY is 
preserved on shell, provided that both electric and magnetic $FI$ terms 
are included. 

\vspace{0.2cm}
\noindent{\bf 4.} The modified $N=2$ SUSY, when 
realized on the superfield strengths (eqs. \p{B9}, \p{modtraV}),  
still closes on space-time translations.
One can wonder how it is 
realized on the gauge-variant objects: gauge potentials and
prepotentials. The $N=2$ SUSY algebra 
itself proves to be modified in this case. We will demonstrate this 
in the $N=1$ superfield formalism, on the example of the model 
considered in Sect. 3.

We define 
\be
W = W_V + \hat{x}^{ik}(\theta_i\theta_k) ~,  \quad  <W_V>_0 = 0\;.
\ee
Decomposing $W_V$ in powers of $\theta_2, \bar \theta^2 $ 
\be
W_V = \varphi (x,\theta_{\s 1}) + i\theta_{\s 2}^\alpha
W_\alpha (x,\theta_{\s 1})
+ (\theta_{\s 2})^2 (1/4)(\bar{D}_{\s 1})^2\bar{\varphi}~,
\label{D10a}
\ee
it is easy to find how the $N=1$ superfield components 
$\varphi (x,\theta_{\s 1})$, 
$W^\alpha (x,\theta_{\s 1}) = (\bar D_{\s 1})^2 D^{\alpha 1}V$ 
behave under the modified $N=2$ SUSY transformations \p{modtraV}. 
We will be interested in how the latter are realized on the 
$N=1$ gauge prepotential $V(x,\theta_{\s 1}, \bar \theta^{\s 1})$.
Modulo gauge transformations, the $N=2$ SUSY acts on $V$ 
as (we choose the $SU(2)$ frame so that 
$M^{12} = 0, M^{11} = M^{22} = m$) 
\be
\delta V = m(\bar{\theta}^{\s 1})^2 \theta^\alpha_{\s 1}
\epsilon_{\alpha\s 2}
+(i/2)\theta^\alpha_{\s1} \epsilon_{\alpha\s 2}\bar{\varphi} +
\mbox{c.c}
+i(\epsilon_{\s 1}Q^{\s 1}+\bar{\epsilon}^{\s 1}\bar{Q}_{\s 1}) V~.
\label{modtraV1}
\ee

Defining $Q_\alpha \equiv Q^1_\alpha $, $S_\alpha \equiv Q^2_\alpha $, 
one finds that the (anti)commutation relations of the $N=2$ SUSY 
algebra are modified as follows: 
\be
\{Q_\alpha, S_\beta \} = \varepsilon_{\alpha\beta} G~, 
\quad \{Q_\alpha, \bar{S}_{\dot{\beta}} \}
= G_{\alpha\dot\beta}~, \quad (\;\mbox{and c.c.}\;)~.  \label{modQS}
\ee
The newly introduced generators possess the following action on $V$
\be 
GV = (i/2)m(\bar \theta^1)^2~, \;\; \bar GV = 
(i/2)m(\theta^1)^2~,\;\; G_{\alpha \dot \beta} V = im\theta_{1\alpha} 
\bar \theta^1_{\dot \beta}~, \label{newgen}
\ee
and are easily seen to be particular $N=1$ gauge transformations. 
Thus, off-shell 
$N=2$ SUSY algebra turns out to be non-trivially unified with the 
gauge group algebra. Such a unification does not contradict 
the famous Coleman-Mandula theorem. Indeed, in gauge theories 
it is impossible to simultaneously satisfy two important 
conditions of this theorem: manifest Lorentz covariance and 
positive definiteness of the metric in the space of states. Note that  
the $N=2$ transformation \p{modtraV1} is defined up to pure 
gauge terms which are capable to change 
the above algebra. However, in the closure of spinor generators
one will always find some gauge generators in parallel 
with $4$-translations. An interesting problem is to construct a non-linear 
realization of this modified $N=2$ algebra along the lines of ref. 
\cite{BG} and to compare it with that constructed 
in \cite{BG} based on the standard $N=2$ algebra. It is also 
tempting to elaborate on a possible stringy origin of the modified 
algebra. 

Finally, we would like to point out that 
the difficulty with the self-consistent incorporation of charged matter 
hypermultiplets into the $N=2$ Maxwell theory with the two sorts of 
$FI$ terms recently discussed in \cite{PP} is directly related to 
the modification of $N=2$ SUSY algebra in the presence 
of these terms. The standard minimal coupling of the $q^+$ hypermultiplets 
to the harmonic superspace $N=2$ Maxwell potential $V^{\s ++}$ \cite{GIK1} 
is not invariant under the modified $N=2$ SUSY. 

\vspace{0.2cm}
\noindent{\bf Acknowledgement.} We are grateful to J. Bagger, 
E. Buchbinder, A. Galperin, S. Krivonos, 
O. Lechtenfeld, D. L\"{u}st, A. Pashnev and M. Vasiliev for useful
discussions. This work was partially supported by 
the grants RFBR-96-02-17634, RFBR-DFG-96-02-00180, 
INTAS-93-127-ext and INTAS-96-308. B.Z. is grateful to Uzbek 
Foundation of Basic Research for the partial support
( grant N 11/97 ).

\end{document}